\documentclass[12pt,preprint]{aastex}
\shortauthors{Reiners et al.}

\begin{document}

\newcommand{\twom}{2MASS\,1207-3932}


\title{Surprisingly Weak Magnetism on Young Accreting Brown Dwarfs}
%{Weak Magnetic Flux on the T~Tauri-like Brown Dwarf
%2MASS~J1207334$-$393254 and Implications for the Scaling of Magnetic
% Fields in T~Tauri Stars}


\author{A. Reiners\altaffilmark{*}}
\affil{Institut f\"ur Astrophysik, Georg-August-Universit\"at,
  Friedrich-Hund-Platz 1, 37077 G\"ottingen, Germany}
\email{Ansgar.Reiners@phys.uni-goettingen.de}
\and
\author{G. Basri}
\affil{Astronomy Department, University of California, Berkeley, CA
  94720 }
\email{basri@berkeley.edu}
\and
\author{U.R. Christensen}
\affil{Max-Planck Institute for Solar System Research,
  Max-Planck-Strasse 2, 37191 Katlenburg-Lindau, Germany}
\email{christensen@mps.mpg.de}


\altaffiltext{*}{Emmy Noether Fellow}


\begin{abstract}
  We have measured the surface magnetic flux on four accreting young
  brown dwarfs and one non-accreting young very low-mass star
  utilizing high resolution spectra of absorption lines of the FeH
  molecule. A magnetic field of 1-2 kG had been proposed for one of
  the brown dwarfs, 2MASS~J1207334$-$393254, because of its
  similarities to higher mass T Tauri stars as manifested in accretion
  and the presence of a jet. We do not find clear evidence for a
  kilo-Gauss field in any of our young brown dwarfs but do find a 2 kG
  field on the young VLM star. Our 3-$\sigma$ upper limit for the
  magnetic flux in 2MASS~J1207334$-$393254 just reaches 1 kG. We
  estimate the magnetic field required for accretion in young brown
  dwarfs given the observed rotations, and find that fields of only a
  few hundred Gauss are sufficient for magnetospheric accretion. This
  predicted value is less than our observed upper limit. We conclude
  that magnetic fields in young brown dwarfs are a factor of five or
  more lower than in young stars of about one solar mass, and in older
  stars with spectral types similar to our young brown dwarfs. It is
  interesting that, during the first few million years, the fields
  scale down with mass in line with what is needed for magnetospheric
  accretion, yet no such scaling is observed at later ages within the
  same effective temperature range. This scaling is opposite to the
  trend in rotation, with shorter rotation periods for very young
  accreting brown dwarfs compared with accreting solar-mass objects
  (and very low Rossby numbers in all cases). We speculate that in
  young objects a deeper intrinsic connection may exist between
  magnetospheric accretion and magnetic field strength, or that
  magnetic field generation in brown dwarfs may be less efficient than
  in stars.  Neither of these currently have an easy physical
  explanation.
\end{abstract}



%% subject header.  Objects should be in the appropriate "individual"
%% headers (e.g. quasars: individual, stars: individual, etc.) with the
%% additional provision that the total number of headers, including each
%% individual object, not exceed six.  The \objectname{} macro, and its
%% alias \object{}, is used to mark each object.  The macro takes the object
%% name as its primary argument.  This name will appear in the paper
%% and serve as the link's anchor in the electronic edition if the name
%% is recognized by the data centers.  The macro also takes an optional
%% argument in parentheses in cases where the data center identification
%% differs from what is to be printed in the paper.


\section{Introduction}

Magnetic fields play an important role in astrophysics; in stars they
regulate accretion of matter onto a star from the very beginning.
Later they are the drivers of magnetic activity and regulate angular
momentum and mass loss throughout the star's lifetime. Young (T~Tauri)
stars can be affected by magnetic fields via two different mechanisms;
(a) magnetospheric accretion, and (b) magnetic activity. Accreting
T~Tauri stars are usually referred to as ``classical'' (CTTS). The
10\% H$\alpha$ line width \citep{White03} is an indicator that can
exclude the accretion scenario on stars with narrower H$\alpha$ lines.
However, in the presence of accretion only sparse information about
magnetic activity can be drawn from emission lines.

Direct measurements of magnetic fields in T~Tauri stars were carried
out, e.g., by \citet{JK99}, \citet{Yang05}, and \citet{JK07}. For
low-mass magnetically active main-sequence stars magnetic fields are
reported in, e.g., \citet{JKV00, Saar01} and \citet{Reiners07}. In
both object classes, typical field strengths are on the order of a few
kilo-Gauss. It is generally believed that the magnetic fields in
active stars are generated through stellar dynamo processes.  In
sun-like stars, the dynamo is probably similar to the solar
(interface) dynamo. In completely convective low-mass stars, however,
a different type of dynamo is required to operate very efficiently
because magnetic flux levels of several kilo-Gauss are apparently
generated ubiquitously in active ultra-cool dwarfs \citep{Reiners07}.
T~Tauri stars are completely convective as well, which implies that
T~Tauri stars could generate magnetic fields through the same dynamo
mechanism operating in low-mass stars.

In the magnetospheric accretion model, accretion is funneled by a
large-scale magnetic field. Gas flows from the disk to the central
star forming an accretion shock and hot spots near the atmosphere of
the accreting objects \citep[e.g.,][]{Koenigl91, Shu94, Hartmann94}.
Outflows in CTTSs are also direcly related to magnetospheric infall
\citep{Hartigan94, Koenigl00}. The strength of the magnetic field
required for this mechanism was calculated by \citet{Koenigl91},
\citet{Cameron93}, and \citet{Shu94}.  According to the scaling
relations derived by these authors, the required surface magnetic
field in an accreting object scales with mass and rotation period
(with weaker magnetic field strength at more rapid rotation).  The
observation that surface rotation velocities in young brown dwarfs are
comparable to those in more massive T~Tauri stars \citep{Mohanty05}
implies that rotation periods in accreting brown dwarfs are
substantially shorter than in higher mass CTTS (because the brown
dwarfs are smaller), therefore the magnetic field required to control
the accretion in the brown dwarfs can be much weaker \citep[see
also][]{JK99}.

For this work, we selected a sample of slowly rotating, young late-M
objects that are bright enough for high resolution spectroscopy.
\cite{Mohanty05} and \cite{Scholz06} provide detailed information on a
large sample of young very-low-mass objects. Our sample consists of
five objects that are members of different star-forming regions with
ages between 1 and 10\,Myrs. Four of the targets are brown dwarfs. One
target, 2MASSW~J1207334-393254, is of particular interest because of
the detection of accretion and an outflow.

\subsection{2MASSW~J1207334-393254}

2MASSW~J1207334-393254 (hereafter \twom) is a $\sim30$\,M$_{\rm Jup}$
brown dwarf \citep{Chauvin05, Mohanty07}. It is a member of the
$\sim$8\,Myr old TW~Hya association \citep{Mohanty03}, and is
surrounded by a disk \citep{Mohanty07, Riaz07}. \twom\ shows clear
signs of accretion in bright asymmetric H$\alpha$ emission
\citep{Mohanty03}, and rapid line emission variability probably due to
rotation of the disk was found by \citet{Scholz06}. \citet{Whelan07}
discovered a bipolar outflow from \twom\ in [O\,\textsc{i}] emission.
Together, this provides strong arguments that \twom\ is a low-mass
analog to the more massive CTTS.

The accretion observed in \twom\ clearly requires the presence of
large-scale magnetic field. However, the strength of that field is not
clear. A field much weaker than several kilo-Gauss could suffice to
maintain the accretion process and mass outflow.

\section{Data}
\label{sect:data}

\begin{deluxetable}{lccc}
  \tablecaption{\label{tab:observations} Log of observations}
  \tablewidth{0pt}
  \tablehead{\colhead{Name} & \colhead{UTC Date} & \colhead{Exp.Time} & Instrument\\
  &&[s]&}
  \startdata
  2MASS $1207-3932$   & 2006-05-06 &  3600 & Keck/HIRES\\
                      & 2007-09-30 &  9800 & VLT/UVES\tablenotemark{a} \\[2mm]
  $\rho$ Oph--ISO 032 & 2008-05-18 &  4400 & Keck/HIRES\\[2mm]
  USco DENIS 160603   & 2006-05-12 &  5400 & Keck/HIRES\\
                      & 2008-05-17 &  4800 & Keck/HIRES\\[2mm]
  CFHT-BD-Tau 4       & 2008-09-20 &  4200 & Keck/HIRES\\[2mm]
  USco 55             & 2008-05-17 &  3000 & Keck/HIRES\\
                      & 2001-06/07 &  3600 & VLT/UVES\tablenotemark{b}\smallskip
  \enddata
  \tablenotetext{a}{PID 279.C-5026}
  \tablenotetext{b}{PID 067.C-0160}
\end{deluxetable}

Most of the data were taken at the W.M. Keck observatory with the
HIRES spectrograph between 2006 and 2008. Our HIRES setup covers the
wavelength range from below H$\alpha$ (6560\,\AA) up to the molecular
absorption band of FeH around 1\,$\mu$m. We used a slit width of
1.15\,\arcsec achieving a resolving power of about $R = 31\,000$. Data
were cosmic-ray corrected, flatfielded, background subtracted, and
wavelength calibrated using a ThAr spectrum. Data reduction was
carried out using routines from the \texttt{echelle} package within
the ESO/MIDAS distribution.  Fringing is not an issue in spectra taken
with the new HIRES CCD, even in very red spectral regions around
1\,$\mu$m.

For \twom, the bulk of our data was obtained by us through a DDT run
with UVES at the Very Large Telescope, ESO, Chile (PID 279.C-5026).
Total exposure time was split into six exposures. We used a slitwidth
of 1\arcsec, which yields a resolving power of $R \approx 40\,000$.
The detector was operated in 1x2 binning mode. We used the standard
UVES setup centered at 860\,nm in the red arm yielding spectral
coverage from 6640\,\AA\ to 10\,450\,\AA. Data were
background-corrected and 2D-flatfielded in standard fashion using
MIDAS-based reduction routines. Sky emission was removed by
subtracting the mean of two sky spectra extracted above and below the
target spectrum in direction perpendicular to dispersion.

Additional data from the ESO
Archive\footnote{\texttt{http://archive.eso.org}} was used for USco~55
(PID 067.C-0160, PI~Guenther). This data was originally taken for
radial velocity monitoring \citep[see][]{Guenther03} and was split
into six exposures of 10\,min duration each. It covers the same
wavelength region as our UVES data for \twom.  Our log of observations
is given in Table\,\ref{tab:observations}.

\section{Measuring rotation and magnetic flux}

The analysis of our spectra follows the strategy laid out in
\citet{Reiners06, Reiners07} and \citet{Reiners09}. To measure the
projected rotation velocity $v\,\sin{i}$ and the magnetic flux $Bf$ of
our sample stars, we utilize the absorption band of molecular FeH
close to 1\,$\mu$m. We compare our data to spectra of the slowly
rotating M-stars GJ\,1002 (M5.5) and Gl\,873 (M3.5).  The magnetic
flux of Gl\,873 was measured to be $Bf = 3.9\,kG$ \citep[using an
atomic FeI line;][]{JKV00}. In order to match the absorption strength
of the target spectra, the intensity of the FeH absorption lines in
the two comparison spectra is modified according to an optical-depth
scaling \citep[see][]{Reiners06}. Note that in the magnetic flux
measurement of our template star, Gl\,873, $Bf$ is the weighted sum of
several magnetic components used for the fit ($\Sigma Bf$). In our
data, we cannot follow such an approach because (1) the spectral
resolution of our data does not allow the differentiation of
individual magnetic components, and (2) no information on the magnetic
splitting pattern of FeH lines is used.  Thus, with our method we can
only scale the product $Bf$ from the spectrum of Gl\,873. We cannot
provide results for $B$ and $f$ individually.

For the determination of the magnetic flux $Bf$, we concentrate on
relatively small wavelength regions that contain absorption lines
particularly useful for this purpose, i.e. regions that contain some
magnetically sensitive as well as magnetically insensitive lines.  We
determine the magnetic flux of our target stars by comparison of the
spectral regions at 9946.0--9956.0\,\AA\ and 9972.0--9981.0\,\AA. In
Figs.\,\ref{fig:stars1}--\ref{fig:stars3} we show the data and the
quality of our fit in the top panels. The SNR of our data is between
20 and 30 in all targets.

For our analysis, we compare the spectra of young, cool stars and
brown dwarfs to template spectra of older stars. We see no reason why
differences in line formation, mainly differences in log~g, would
affect our measurement because all FeH lines belong to the same
ro-vibrational transition. If FeH line formation is affected by a
difference in log~g, this would apply to \emph{all} spectral lines.
The strength of our differential technique is that we try to match
magnetically sensitive and non-sensitive lines at the same time.
Systematic effects would not result in fake magnetic field
measurements but in an overall degradation of fit quality, which we do
not see. An upper limit in our method implies that the scaled template
of the inactive star did a good job of fitting the target spectrum. A
preliminary study on the behaviour of FeH at different gravities was
carried out by \citet{Wende08} using 3D-hydrodynamical simulations. No
systematic effect on the FeH lines, in particular between groups of
magnetically more or less sensitive lines, was found.

\section{Results}

\begin{deluxetable}{lccccccccccc}
  \tabletypesize{\scriptsize} %only for preprint style 
  \tablecaption{\label{tab:results} Results, some additional
    parameters and estimates on the magnetic flux using different
    assumptions} \tablewidth{0pt}
  \tablehead{\colhead{Name} & Spectral & $v\,\sin{i}$ & accretor\tablenotemark{a} & $Bf$& Age\tablenotemark{b} & $T_{\rm eff}$\tablenotemark{c} & $M$\tablenotemark{d} & $R$\tablenotemark{d} & $\log{L}$\tablenotemark{d} & $Bf_{\rm scale}$ & $B_{\rm funnel}$ \\
    & Type &[km\,s$^{-1}$] & & [kG]
    &[Myr]&[K]&[M$_\odot$]&[R$_\odot$]&[L$_\odot$]&kG&kG} \startdata
  2MASS $1207-3932$   & M8.0 &           16$^{+1}_{-3}$ & Y   & 0.0$^{+1.0}       $ & 10 & 2700 & 0.037 & 0.26 & $-2.48$ & $2.0^{+1.3}_{-1.2}$ & 0.2 \\[1mm]
  $\rho$ Oph--ISO 032 & M8.0 &           12$^{+3}_{-3}$ & Y   & 0.6$^{+1.8}_{-0.6}$ &  1 & 2700 & 0.032 & 0.44 & $-2.10$ & $1.4^{+0.9}_{-0.8}$ & 0.1 \\[1mm]
  UpSco DENIS 160603   & M7.5 &           14$^{+1}_{-1}$ & Y   & 0.0$^{+0.4}       $ &  5 & 2800 & 0.042 & 0.40 & $-2.05$ & $1.7^{+1.1}_{-1.0}$ & 0.1 \\[1mm]
  CFHT-BD-Tau 4       & M7.0 & \phantom{1}6$^{+2}_{-4}$ & Y ? & 0.8$^{+1.0}_{-0.8}$ &  2 & 2900 & 0.064 & 0.65 & $-1.57$ & $1.5^{+1.0}_{-0.9}$ & 0.2 \\[1mm]
  UpSco 55 & M5.5 & \phantom{1}9$^{+4}_{-1}$ & N & 2.3$^{+1.2}_{-1.4}$
  & 5 & 3050 & 0.095 & 0.52 & $-1.68$ & $1.9^{+1.3}_{-1.1}$ & 0.3
  \smallskip
  \enddata
  \tablenotetext{a}{from H$\alpha$ 10\,\% width \citep[see][]{White03}}
  \tablenotetext{b}{see \citet{Mohanty05}}
  \tablenotetext{c}{from spectral type according to \citet{Luhman03}}
  \tablenotetext{d}{from age and $T_{\rm eff}$ according to \citet{Baraffe98}}
\end{deluxetable}

\begin{figure}
  \centering
  \mbox{
    \parbox{0.45\hsize}{
      \mbox{\includegraphics[width=\hsize]{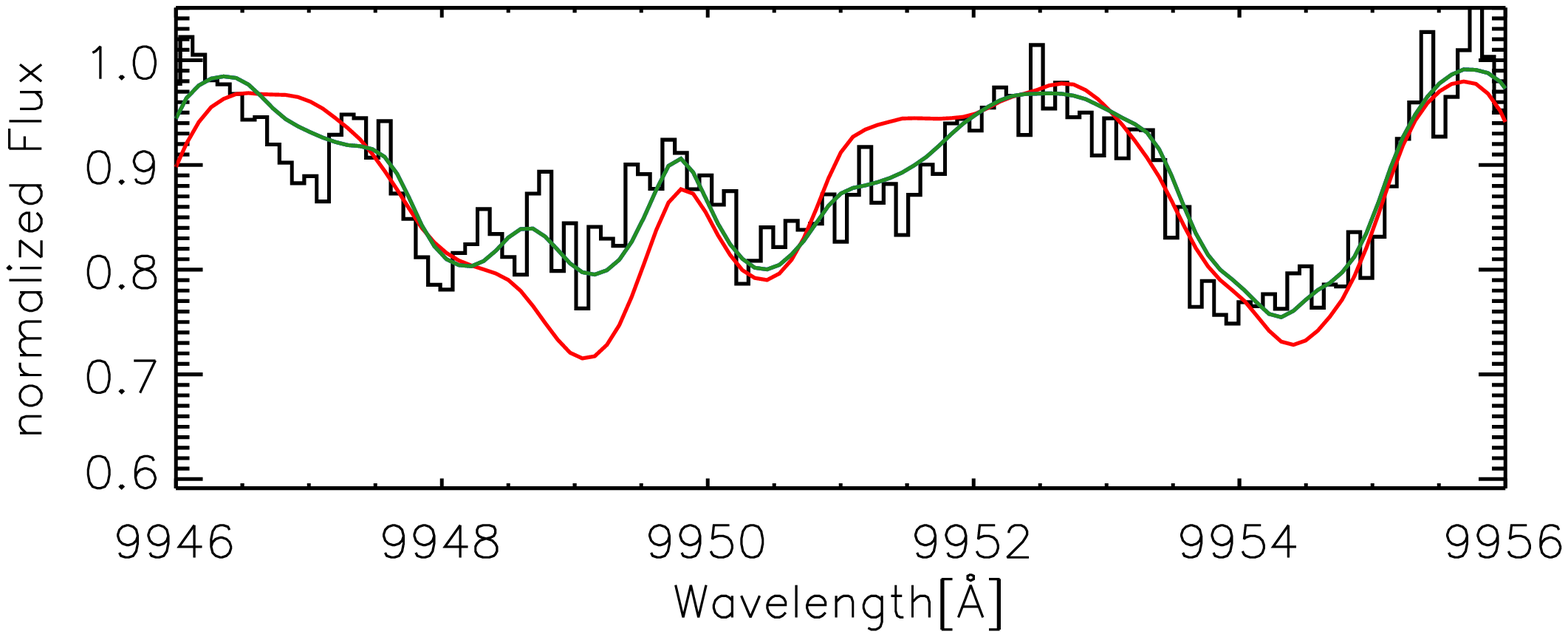}}\\
      \mbox{\includegraphics[width=.97\hsize]{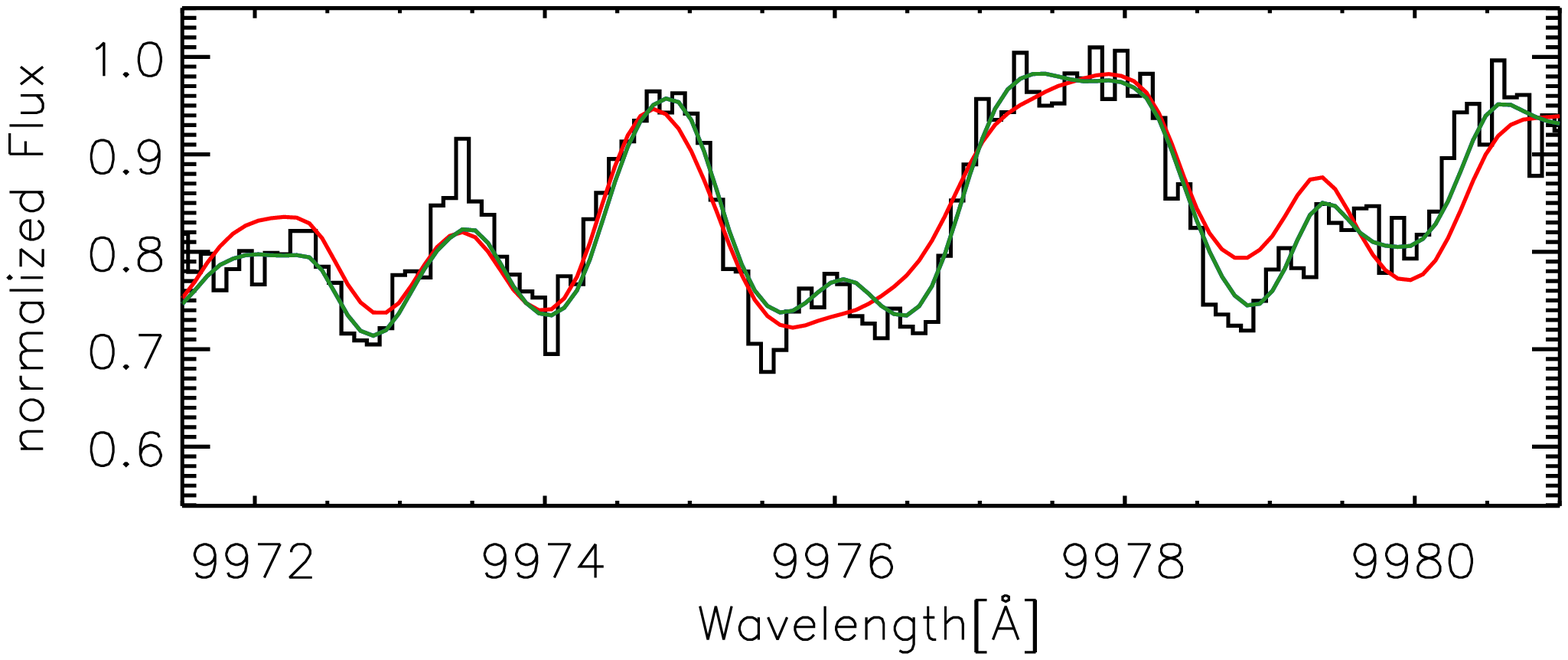}}\\[2mm]
      \mbox{\includegraphics[width=.96\hsize]{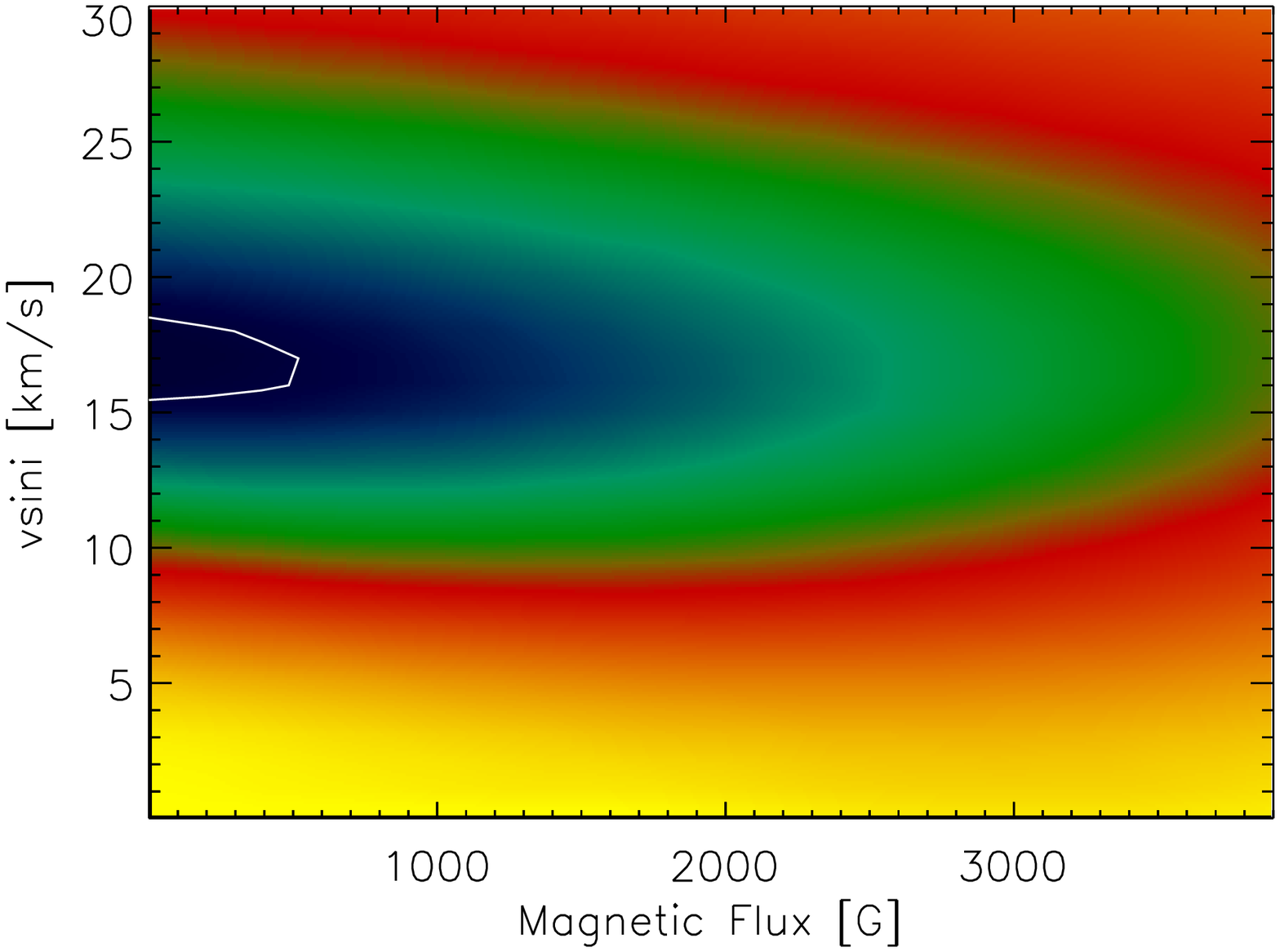}}
    }
    \parbox{.45\hsize}{
      \mbox{\includegraphics[width=\hsize]{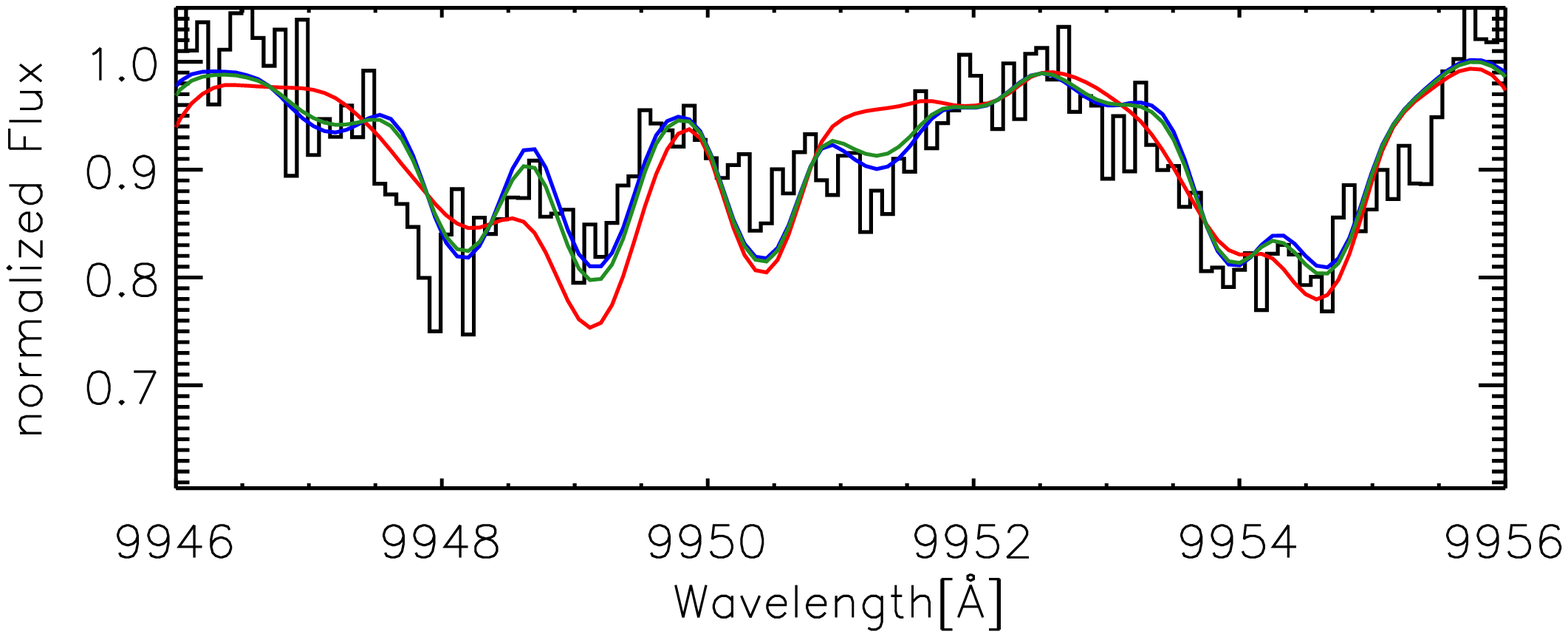}}\\
      \mbox{\includegraphics[width=.97\hsize]{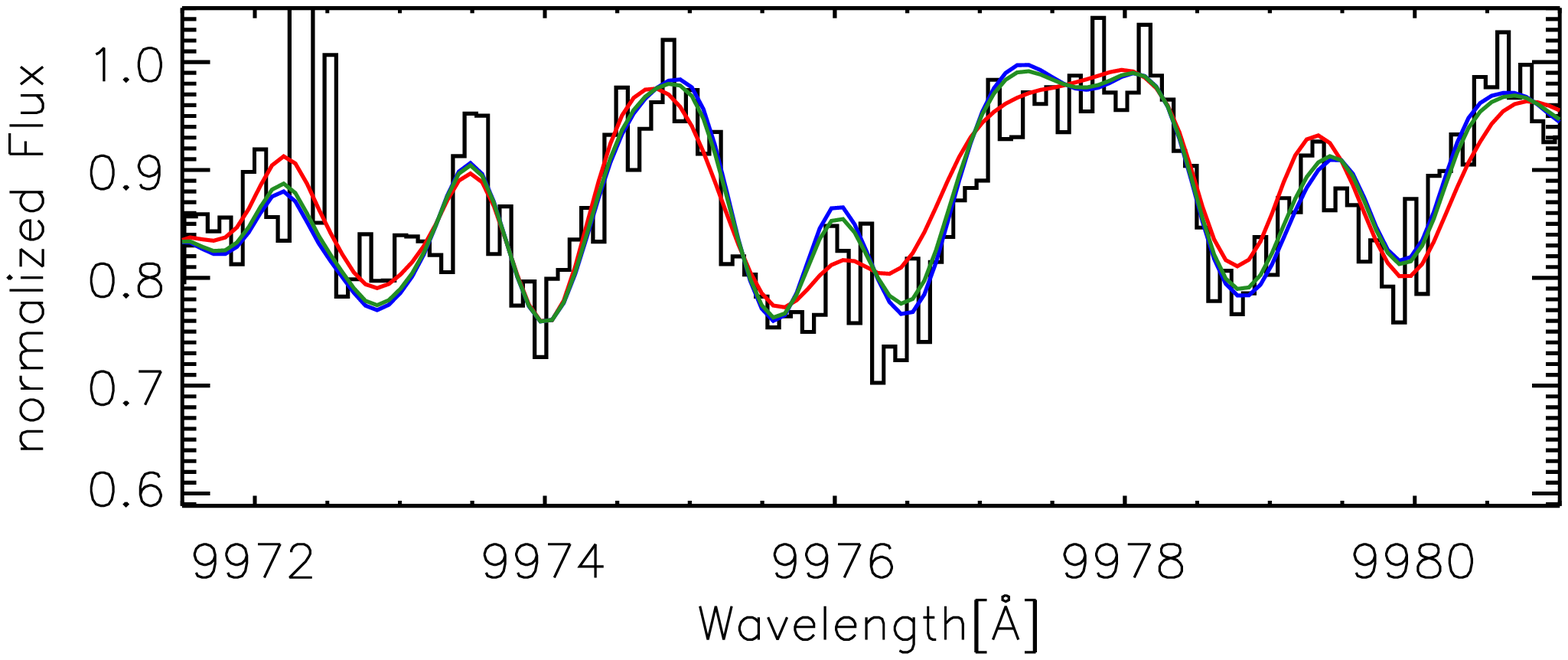}}\\[2mm]
      \mbox{\includegraphics[width=.96\hsize]{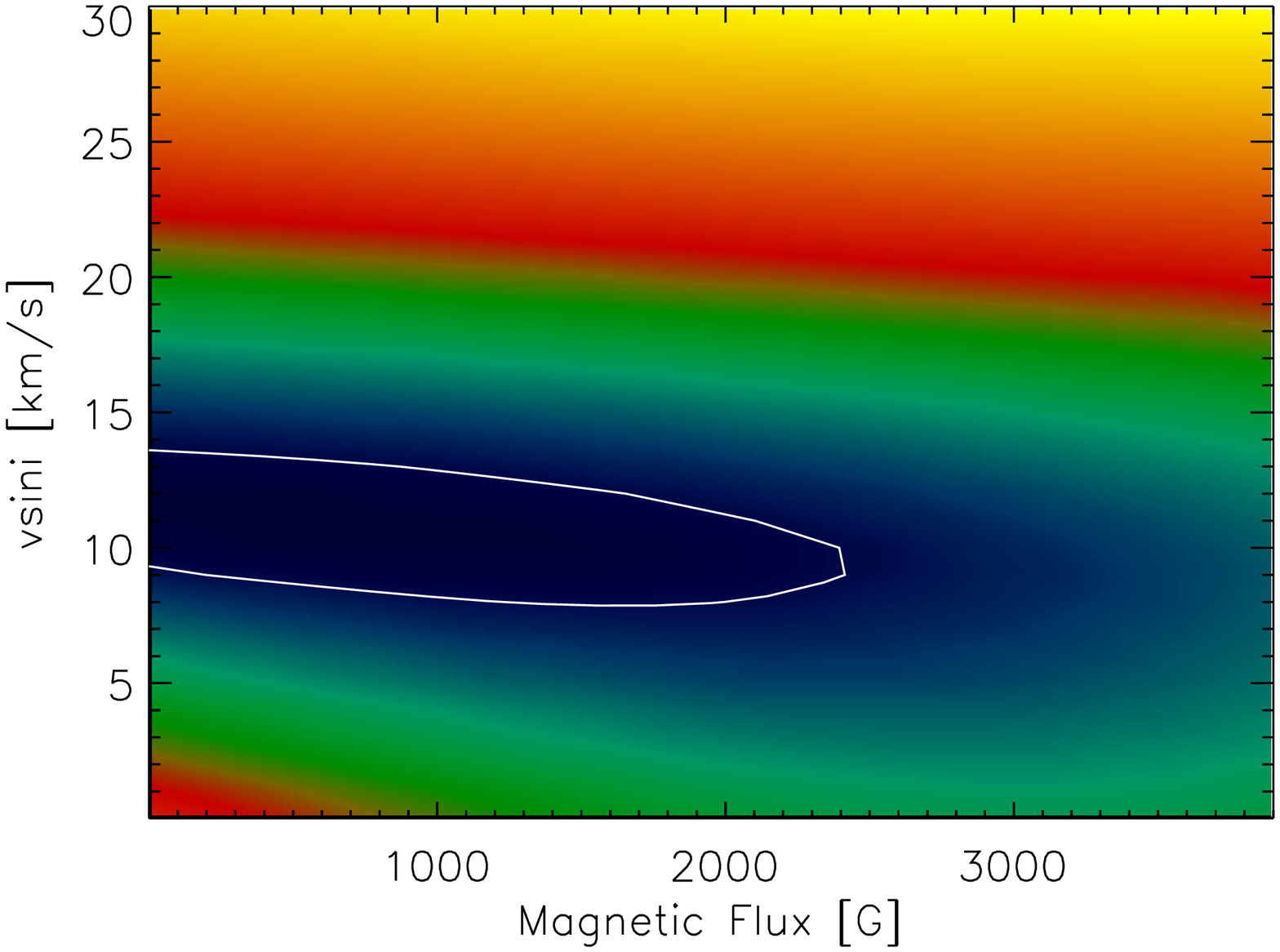}}
    }
  }
  \caption{\label{fig:stars1}Data and fit (top panel) and
    $\chi^2$-landscapes (bottom panel) for 2MASS1207 (left) and
    ISO~032 (right). In the upper two panels, data are shown as black
    histograms. The three coloured lines show our fit for no magnetic
    field (blue line), strong magnetic flux ($Bf \sim 4$\,kG, red
    line), and the best fit, which is an interpolation of the two
    (green line). In some cases, the green line is covering the blue
    line. In the bottom panel, dark and blue colour indicates low
    $\chi^2$ values, red and yellow colours show bad fit quality (high
    values of $\chi^2$. The white line shows the 3$\sigma$-level. }
\end{figure}

\begin{figure}
  \centering
  \mbox{
    \parbox{0.45\hsize}{
      \mbox{\includegraphics[width=\hsize]{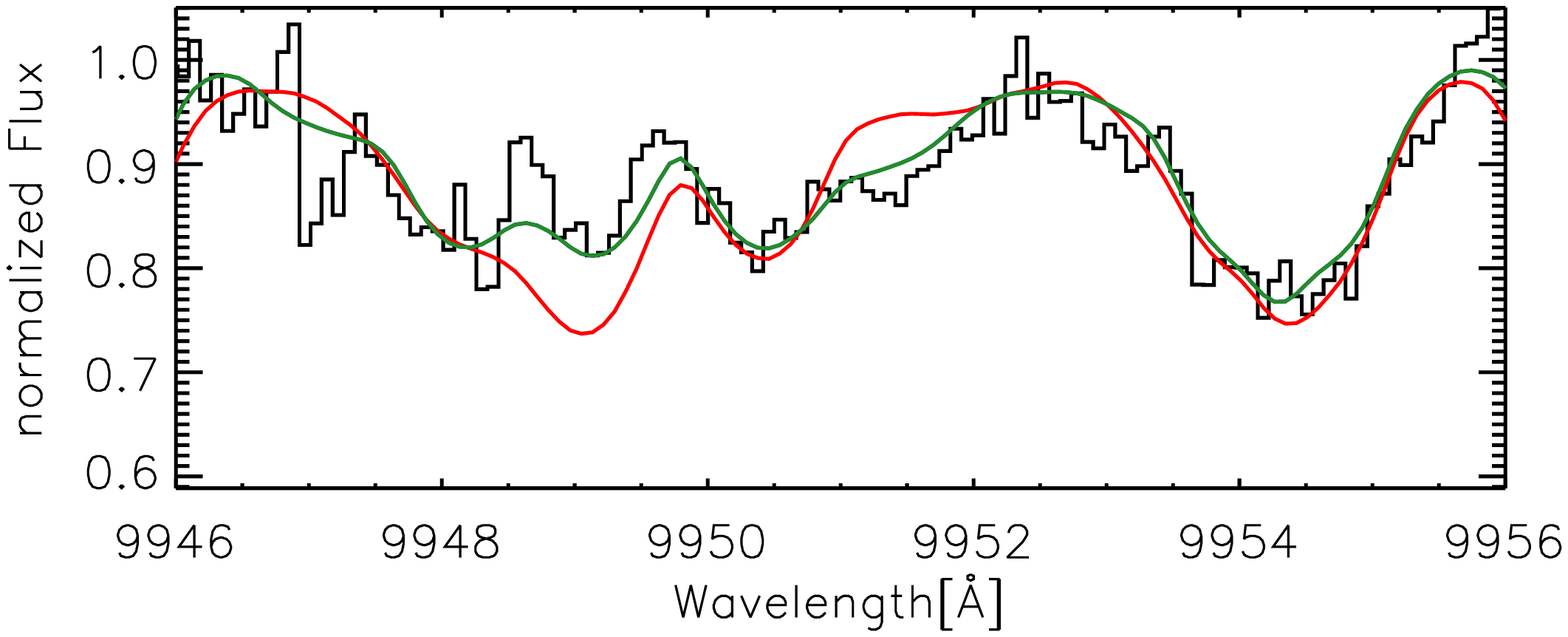}}\\
      \mbox{\includegraphics[width=.97\hsize]{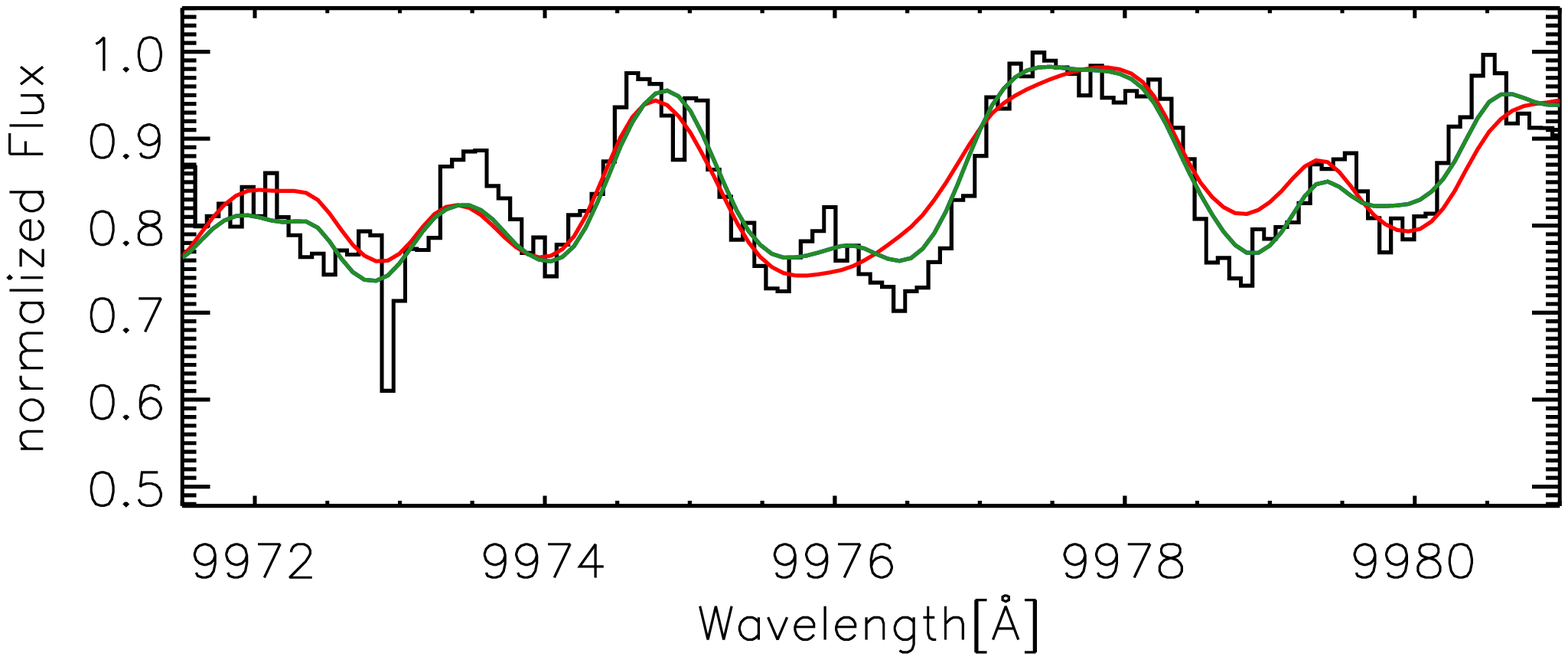}}\\[2mm]
      \mbox{\includegraphics[width=.96\hsize]{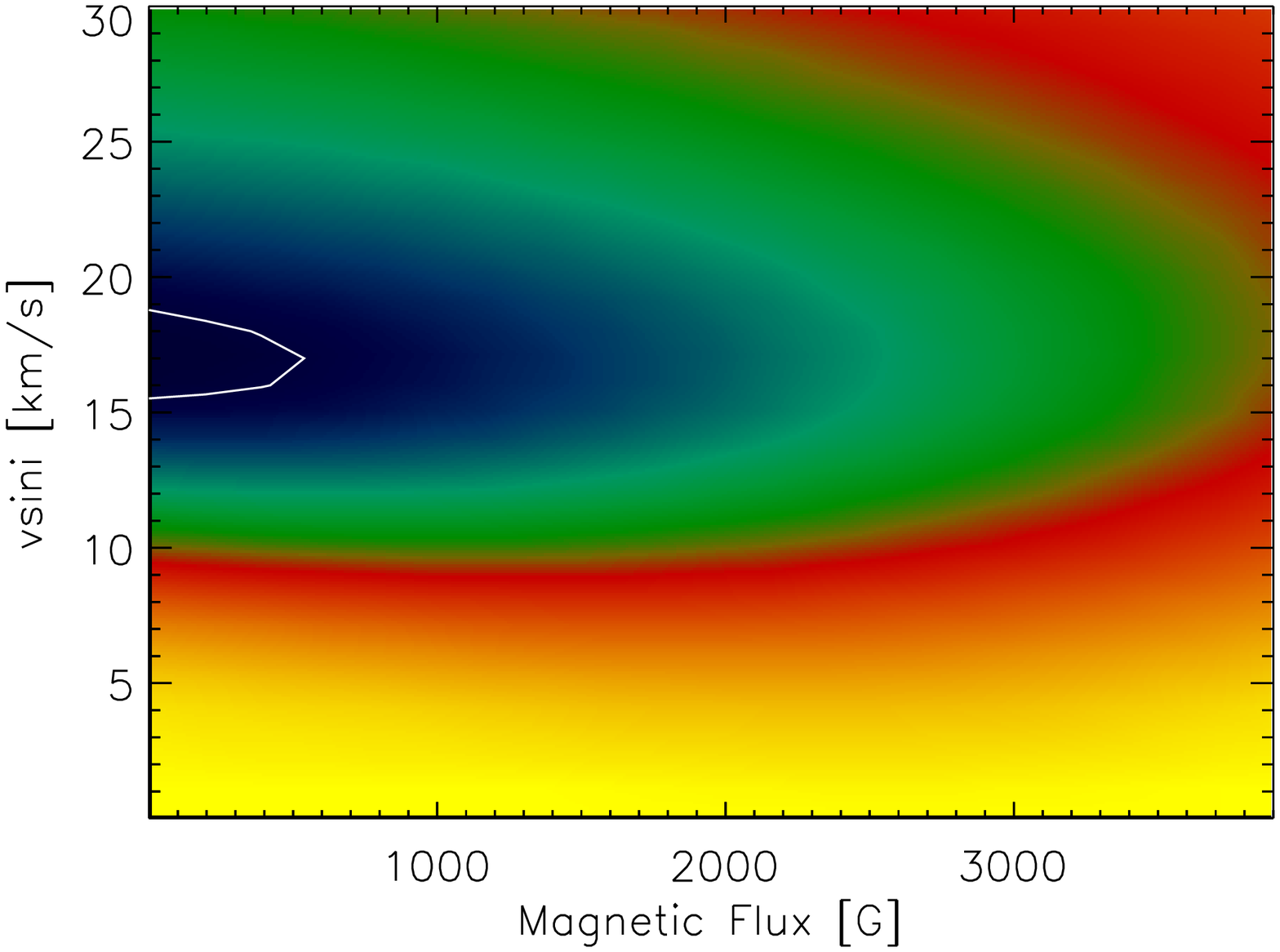}}
    }
    \parbox{0.45\hsize}{
      \mbox{\includegraphics[width=\hsize]{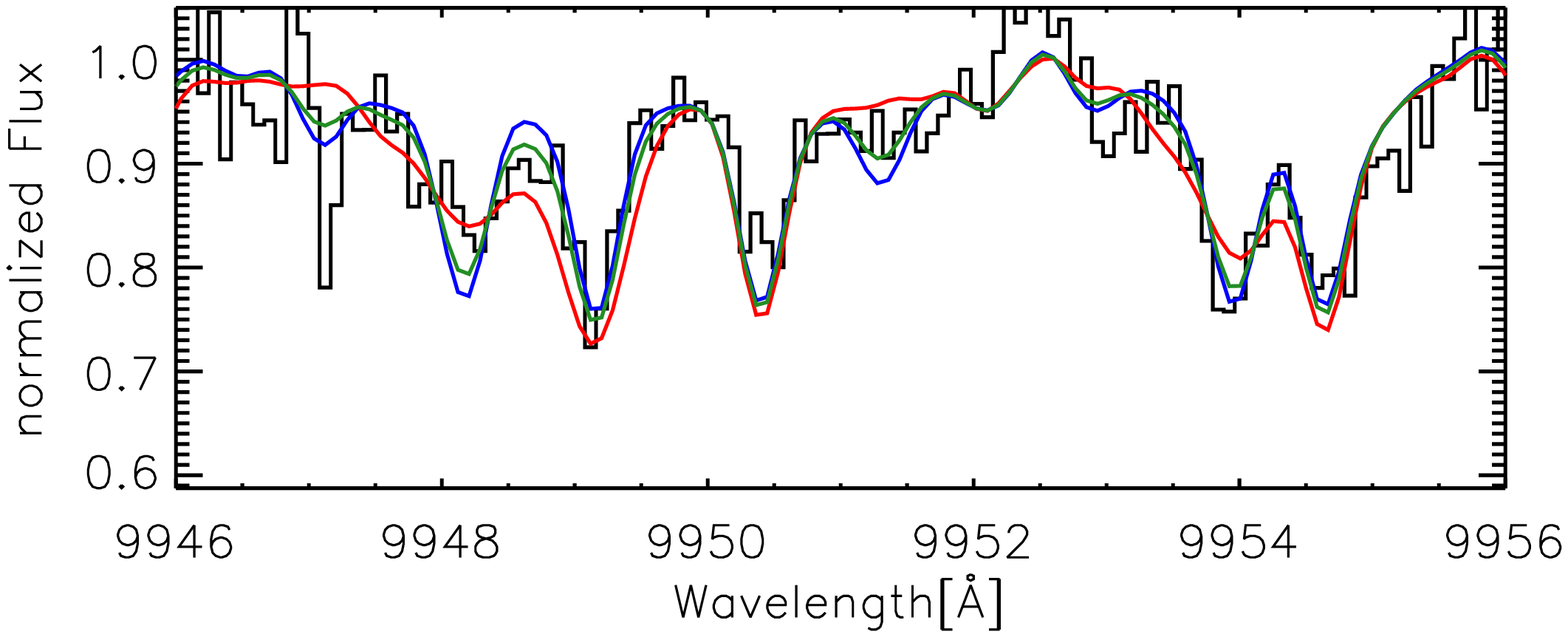}}\\
      \mbox{\includegraphics[width=.97\hsize]{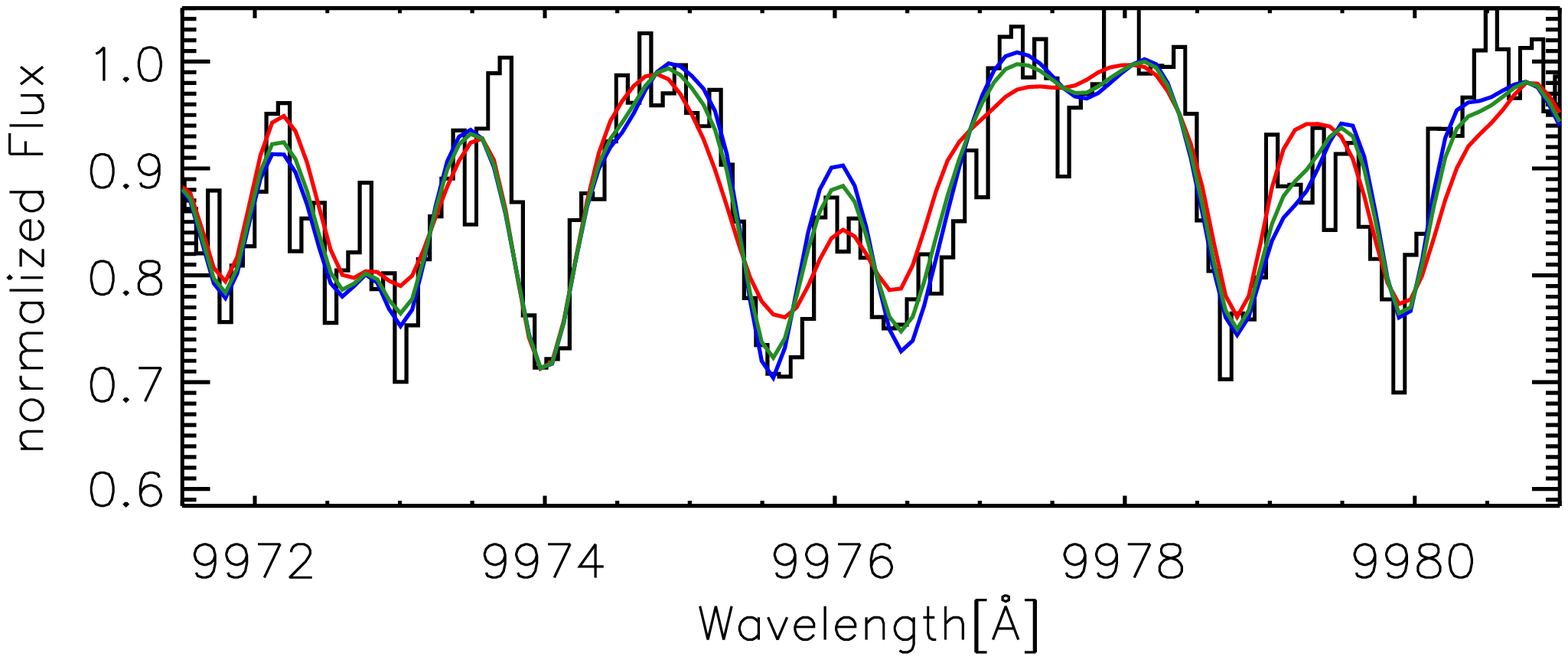}}\\[2mm]
      \mbox{\includegraphics[width=.96\hsize]{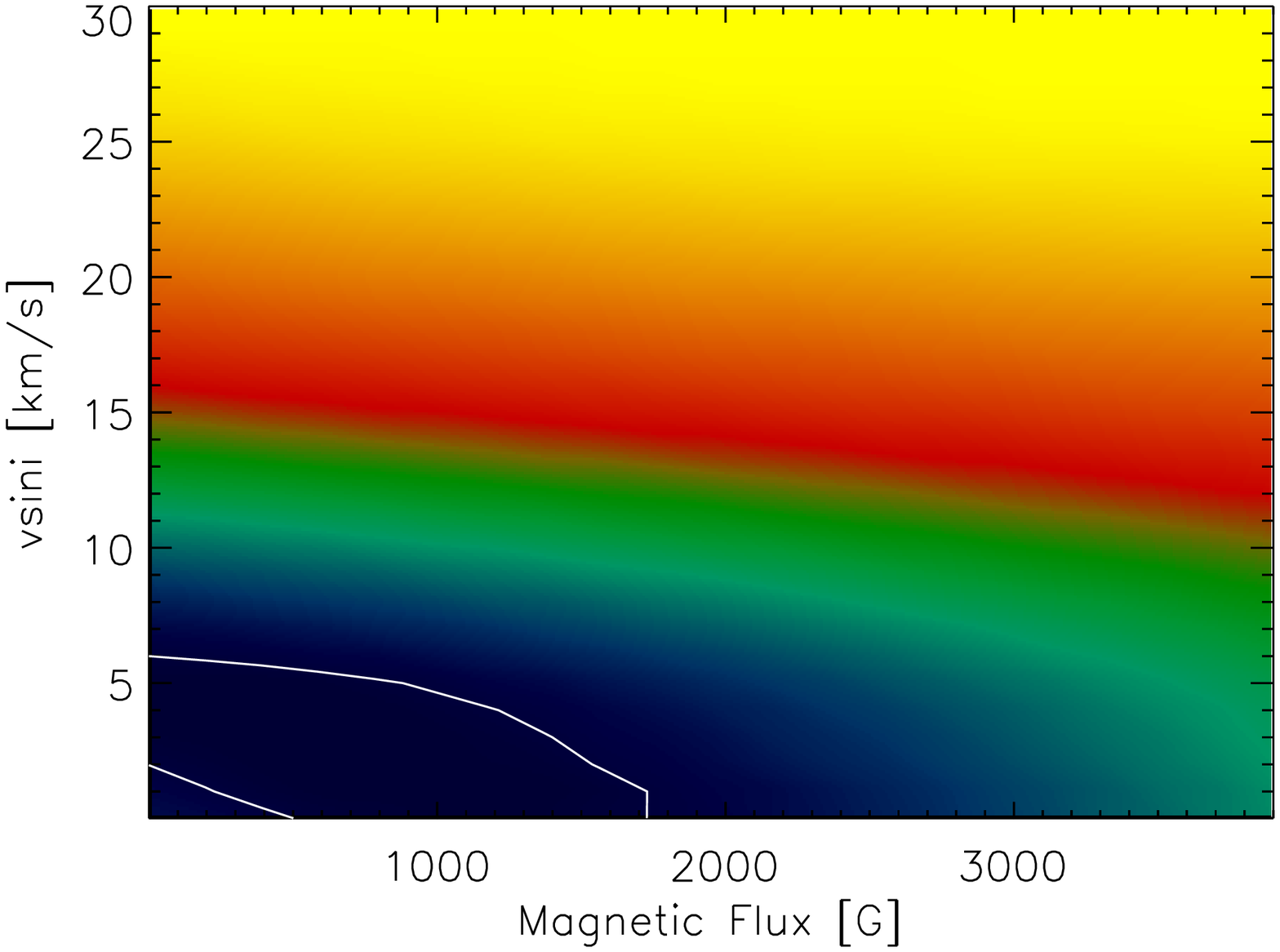}}
    }
  }
  \caption{\label{fig:stars2}Data and fit (top panel) and
    $\chi^2$-landscapes (bottom panel) for UpSco~DENIS~1606 (left)
    and CFHT~4 (right). See Fig.\,\ref{fig:stars1} for more explanation.}
\end{figure}

\begin{figure}
  \centering
  \mbox{
    \parbox{.45\hsize}{
      \mbox{\includegraphics[width=\hsize]{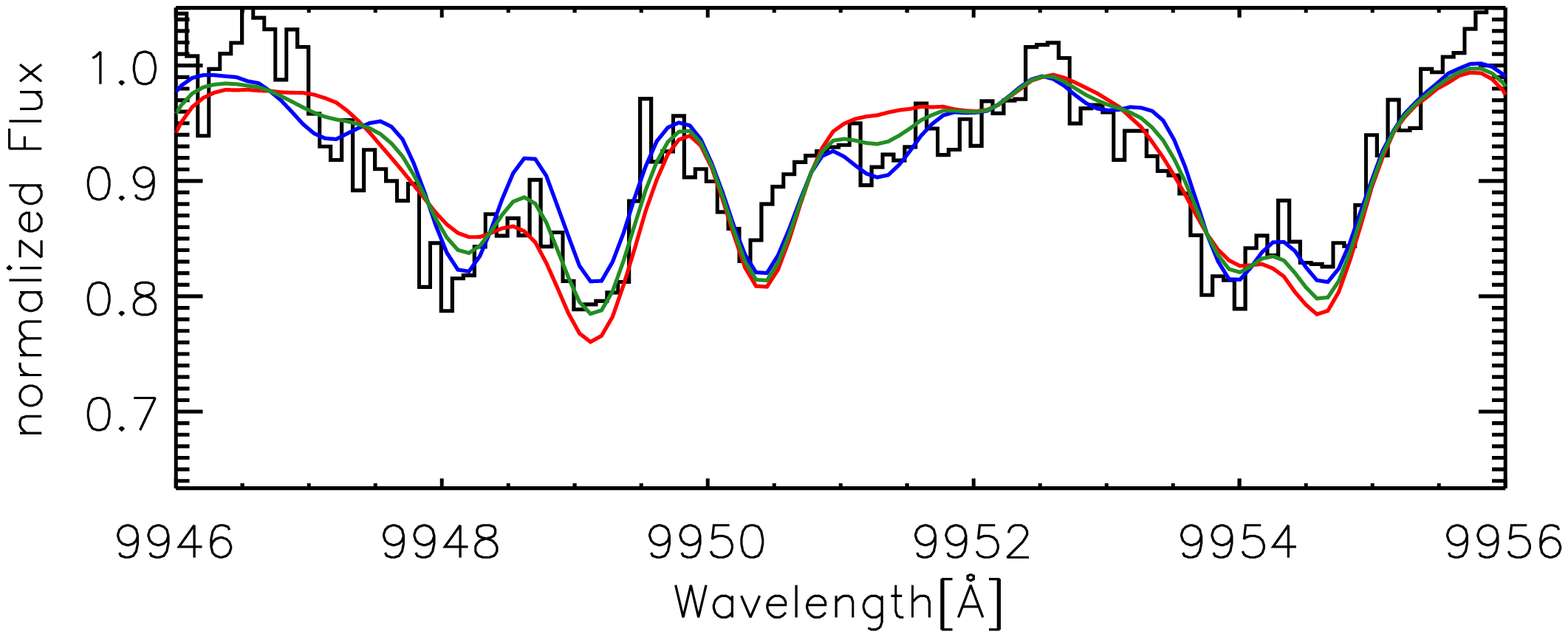}}\\
      \mbox{\includegraphics[width=.97\hsize]{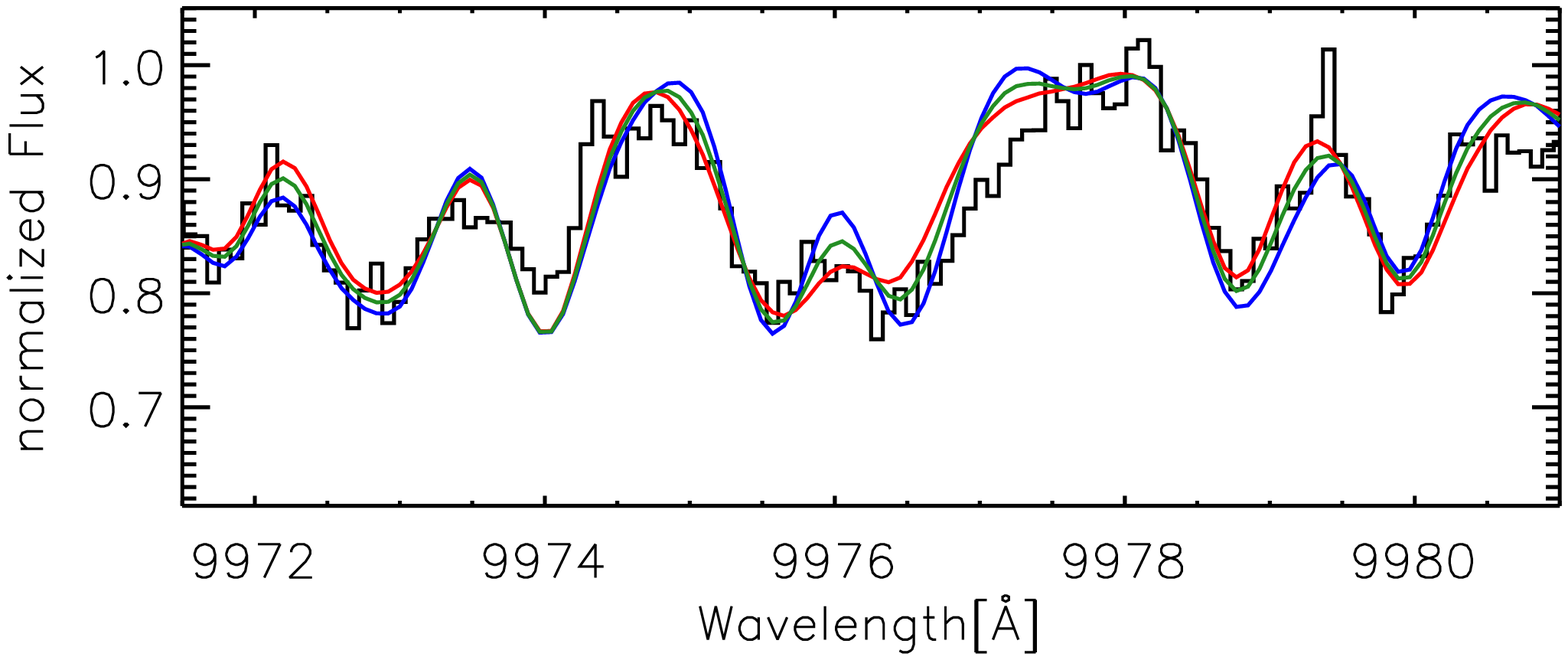}}\\[2mm]
      \mbox{\includegraphics[width=.96\hsize]{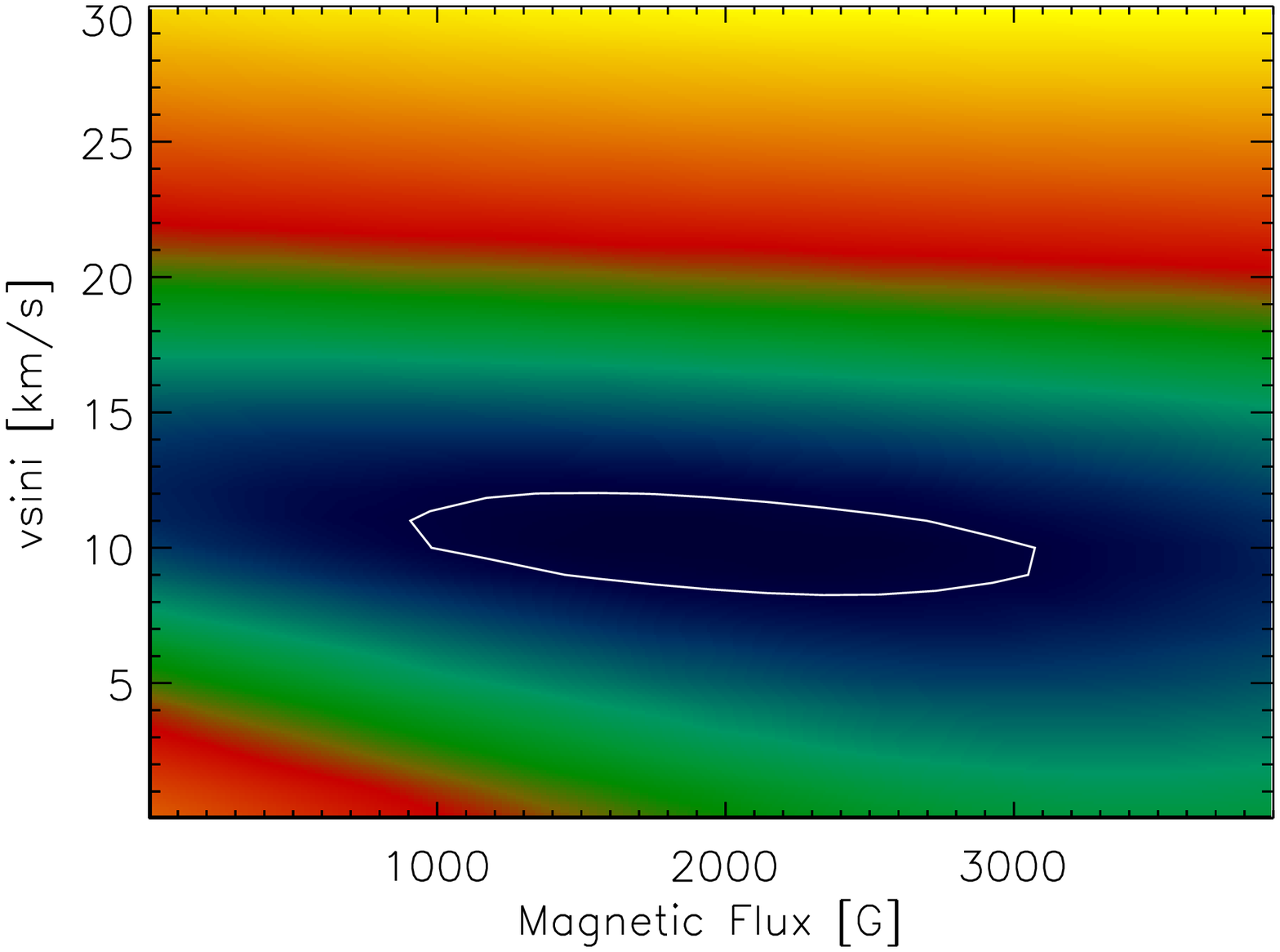}}
    }
  }
  \caption{\label{fig:stars3}Data and fit (top panel) and
    $\chi^2$-landscape (bottom panel) for UpSco~55. See
    Fig.\,\ref{fig:stars1} for more explanation.}
\end{figure}

We show the fits to our data in the upper panels of
Figs.\ref{fig:stars1}--\ref{fig:stars3}. We overplot the two extreme
cases (a) without any magnetic field, and (b) strong magnetic flux
with $Bf \approx 4$\,kG. Template spectra are artificially broadened
and scaled to match the absorption depth of the FeH lines in our
sample targets \citep[see][]{Reiners07}. The lower panels of
Figs.\ref{fig:stars1}--\ref{fig:stars3} show $\chi^2$-maps, i.e., the
goodness of fit, $\chi^2$, as a function of projected rotational
velocity, $v\,\sin{i}$, and total magnetic flux, $Bf$. We mark the
formal 3$\sigma$-region, $\chi^2 = \chi^2_{\rm min} + 9$, with a white
line. In all cases, the values of the reduced $\chi^2$,
$\chi^2_{\nu}$, is on the order of unity. As expected, the
3$\sigma$-regions are usually extended in $Bf$ but relatively narrow
in $v\,\sin{i}$. There is always a small tilt in the shape of these
regions indicating the slight degeneracy between rotation and magnetic
flux because both effects result in a broadening of spectral lines
(but $Bf$ only broadens a subset of all lines).

The results of our analysis for all stars are given in
Table\,\ref{tab:results}. Values of $v\,\sin{i}$ and $Bf$ are best
fits from our $\chi^2$ analysis with formal 3$\sigma$-uncertainties.
The uncertainties in $Bf$ are usually on the order of 1~kG reflecting
the small effect of $Bf$ on the spectral lines and the sometimes
limited quality of our data. With this large an uncertainty we cannot
rule out magnetic flux on the order of 1~kG or lower. Nevertheless, we
are able to distinguish between flux values on a kilo-Gauss scale. In
Table\,\ref{tab:results}, we have also included estimates on stellar
parameters including age, effective temperature, mass, radius, and
bolometric luminosity (see below). Two estimates for magnetic flux
from different perspectives are included as well, these will be
explained in the following.

In only one of our five sample stars do we find magnetic flux on the
kG-level; USco~55 shows magnetic flux of $Bf \approx 2 \pm 1$\,kG.  In
all other objects, only upper limits on the order of 1\,kG are found.
The preferred values for these would be several hundred Gauss; 1kG is
the upper limit and it is unlikely that all objects would be at this
three sigma maximum. In $\rho$~Oph-ISO~032, data quality is lower and
the upper limit exceeds 2\,kG.

\section{Discussion}

Although the uncertainties in magnetic flux measurements in these
faint objects are quite large, the magnetic field detection in
UpSco~55 stands in marked contrast to the non-detections in the other
four targets. Magnetic flux on the 2\,kG level can be excluded for
most of the four, and it is rather unlikely in all. In the following,
we discuss implications and some possible explanations for this
result.

The low-mass objects without magnetic flux detections represent the
first objects of spectral type M6 and later in which magnetic flux is
found to be weaker than a kilo-Gauss.  Eight objects of spectral type
M6 and later were investigated by \citet{Reiners07}, and all of them
show strong magnetic flux. The main difference between our targets and
the objects in \citet{Reiners07} is their age, implying that they are
of different mass. The four weakly magnetic objects of our sample can
be considered brown dwarfs with quite some confidence, i.e., the mass
estimates from age and spectral type are well below the theoretical
threshold of hydrogen burning ($M = 0.08$\,M$_\odot$). UpSco~55, on
the other hand, is well above that boundary and hence is not a brown
dwarf. In contrast to young brown dwarfs of spectral type late-M, a
sample of late-M field stars at an average age of a few Gyrs
\citep[like in][]{Reiners07} have masses above the hydrogen burning
limit as well.  Hence, all other late-M objects with detected strong
magnetic flux are not brown dwarfs; this is confirmed by the fact they
all fail the lithium test \citep{Basri98}.

\subsection{Magnetospheric Accretion}

The second attribute that distinguishes these four brown dwarfs from
older stars, and from UpSco~55, is the presence of magnetospheric
accretion. A significant difference between young and old objects is
that young objects are still surrounded by a disk to which they may be
coupled through their magnetic field. The accretion of disk material
in CTTS is believed to be governed by a magnetic field intercepting
the disk and channeling the accretion flow onto the stellar surface.
\citet{Koenigl91} provided the classic relation how the magnetic field
is related to stellar mass and radius, accretion rate, and angular
velocity (see his Eq.\,3).  This relation is also connected to the
inner radius of the disk, i.e.  the radius where the disk is
truncated. Using representative values for a model T~Tauri star
\cite[see][]{Koenigl91} with $M = 0.8$\,M$_\sun$, \textit{\.M} =
$10^{-7}$\,M$_\sun$\,yr$^{-1}$, $R = 2.5$\,R$_\sun$, and $\Omega = 1.7
\times 10^{-5}$\,s$^{-1}$, \citet{Koenigl91} predicts a magnetic field
of kilo-Gauss strength on the surface of the star. The inner edge of
the disk is then at a radius of $R_{\rm in} = 2$\,R$_*$. Stellar
magnetic fields of kilo-Gauss strength are indeed found in a number of
T~Tauri stars indicating that the model of \cite{Koenigl91} predicts
field values of realistic magnitude. In fact, most TTS are found to
have disks truncated further out than that, and even stronger surface
fields \citep[see e.g.][]{JK99, Yang05, JK07}. \citet{Cameron93} and
\citet{Shu94} derived similar relations for magnetic fields in
accreting stars.  While \citet{Cameron93} predict magnetic fields of
comparable strength, the estimate of \citet{Shu94} predicts magnetic
fields about a factor of two weaker \citep[cf.  ][]{JK99}.

Using the equation of \citet{Koenigl91}, \citet{Scholz06} and
\citet{Stelzer07} predict the surface field strength in brown dwarfs
to be about half as strong as in CTTS, which would be still in the
kilo-Gauss regime.  These authors make the assumption that the inner
disk radius is constant at $R_{\rm in} = 2$\,R$_*$. This determines
the angular rotation velocity of the objects. In fact, this assumption
implies a surface rotation rate of $v = 40$\,km\,s$^{-1}$ for their
model star.  Interestingly, from their calculation, \citet{Stelzer07}
predict a magnetic field of only $\sim 200$\,G for \twom.  However,
for a more massive star they also derive a value of only $\sim 600$\,G
($M = 0.8$\,M$_\sun$, \textit{\.M} = $10^{-8}$\,M$_\sun$\,yr$^{-1}$,
$R = 1.5$\,R$_\sun$), which is substantially lower than the
multikilo-Gauss fields found in CTTS. \citet{Stelzer07} speculate that
considerable uncertainties lead to this inconsistency. They suggest
that both results, the 200\,G for the brown dwarf and the 600\,G for
CTTS, underestimate the real fields by some factor. According to this
argumentation, brown dwarfs should have fields a factor 2--3 weaker
than CTTS, i.e. still on the order of a kilo-Gauss.

Instead of assuming a global underestimate of the field strengths, we
can also estimate the magnetic field without the assumption of
constant truncation radius, but assuming that the higher-mass star
rotates at a lower surface velocity of only $v = 10$\,km\,s$^{-1}$ (as
is typically observed). In this case, one finds a surface magnetic
field of $B \sim 2.8$\,kG for CTTS, which is more in line with
observations. In other words, a somewhat slower rotation (hence larger
truncation radius) easily pushes the surface magnetic field of a CTTS
into the kilo-Gauss regime.

We can now apply the equations of \citet{Koenigl91} to our sample
objects, without the assumption of a constant truncation radius. The
lower limits of the measured surface rotation velocities (i.e., $i =
90\degr$) lead to upper limits of the magnetic fields of $B_{\rm
  funnel} \sim $100--300\,G.  They are given in
Table\,\ref{tab:results}. Note that these are upper limits -- $B_{\rm
  funnel}$ would be even lower if the stars are viewed under
inclination angles $i < 90\degr$ implying more rapid rotation.  This
result is fully consistent with the non-detection of magnetic fields
from our data.  We conclude from this that our upper limits of
magnetic flux in accreting brown dwarfs are not inconsistent with the
magnetospheric accretion model. The ratio between surface magnetic
flux necessary for magnetospheric accretion in CTTS of nearly one
solar mass and in young accreting brown dwarfs is on the order of ten
rather than two, given the observed values of surface rotation. There
is an extensive literature of rotations for CTTS, and a good initial
set of measurements for young brown dwarfs can be found in
\citet{Mohanty05}.

While the weak magnetic fields are consistent with magnetospheric
accretion, this does not explain the weakness of the fields, because
magnetospheric accretion is thought to be a consequence of the
presence of a magnetic field rather than a process that determines the
magnetic field strength. Current models also assume that accretion
applies no net torque on the object \citep{GLa, GLb}. Hence, it is not
clear how accretion could explain that young brown dwarfs have fields
much weaker than their older stars of the same spectral type.

\subsection{Magnetic Flux Scaling with Energy Flux}

As mentioned above, the four late objects of our sample represent the
first objects later than spectral type M5.5 on which no magnetic flux
is detected. We can speculate why the magnetic field is rather weak on
the surface of young brown dwarfs.  A reasonable explanation might
involve the fact that young brown dwarfs have not yet contracted,
since the deuterium burning phase for these objects lasts about 10
Myr. Perhaps the magnetic field grows stronger as the star later
shrinks.  \citet{Christensen09} presented a model in which the
available energy flux determines the magnetic field strength of
rapidly rotating objects (a few km\,s$^{-1}$).  This model is
successful in estimating magnetic fields in planets of our solar
system, in field stars, and in T~Tauri stars. One prediction of the
model is that low-mass objects have weaker fields than high-mass
objects, which is the very trend we observe in our small sample. We
tested the hypothesis that energy flux of the objects explains the
difference we see in magnetic flux \citep[for more details,
see][]{Christensen09}. From the age and spectral type of the targets
we estimate their mass, radius, and bolometric luminosity.  We follow
the conversion from spectral type to effective temperature by
\citet{Luhman03} and convert temperature and age into the other
parameters according to the models of \citet{Baraffe98}. The predicted
flux expressed in terms of mass, radius, and luminosity is

\begin{equation}
  Bf_{\rm scale} \sim \left(\frac{M}{M_{\sun}}\right)^{1/6} \left(\frac{L}{L_{\sun}}\right)^{1/3}
  \left(\frac{R}{R_{\sun}}\right)^{-7/6} \times 4.8^{+3.2}_{-2.8}\,{\rm kG},
\end{equation}
where the uncertainties are estimated along the lines discussed in
\citet[][online supplementary information]{Christensen09}.

The magnetic flux predicted from an energy flux scaling, $Bf_{\rm
  scale}$, are given in Table\,\ref{tab:results}. They are in the
range 1.4--2.0\,kG for all five targets including UpSco~55, for which
a value of 1.9\,kG is predicted. The only object in which we find
magnetic flux consistent with the expectation from an energy flux
scaling is UpSco~55. According to this scaling, the differences in
magnetic flux between UpSco~55 and the brown dwarfs should not exceed
a few tens of percent.

\subsection{Constraints from Rotation and Activity}

There is also increasing evidence that even fully convective dynamos
are sensitive to rotation period \citep{Reiners07}. A potential way to
explain the weak magnetic fields found in our young sample could be
slow rotation. Observations have shown that in older brown dwarfs the
rotation period decreases down to hours \citep{MohantyB03, Zapatero06,
  Reiners08}. Thus, if magnetic flux generation is not saturated in a
brown dwarf, it is likely to increase with age and might be rather low
in young brown dwarfs. On the other hand, magnetic field generation
obviously is effective in UpSco~55, which has a rather similar
rotation period to the others. Its surface temperature and luminosity
are rather similar to CFHT~4 (because that object is both younger and
less massive). It is also a pre-main sequence object (unlike the stars
in the older sample) and so has a similar interior structure and
energy source compared to the brown dwarfs.  Furthermore, the rotation
periods of our targets are in the range of 1--6\,d so that Rossby
numbers are probably on the order of $\sim10^{-2}$.  This is well in
the regime of saturated flux generation found in old M dwarfs
\citep{Reiners09}.  Thus, slow rotation is a rather unlikely
explanation for the lack of magnetic flux in young brown dwarfs.

Signs of magnetic activity are seen on both young accreting and
non-accreting brown dwarfs, so the magnetic fields are strong enough
to produce atmospheric heating in all cases.  \citet{Grosso07} find
that young brown dwarfs achieve fractional X-ray luminosities that are
about a factor of four smaller than for older main-sequence stars of
similar effective temperature.  These authors find no evidence for a
difference in X-ray emission between accreting and non-accreting brown
dwarfs, although CTTS tend to show less X-ray activity than
non-accreting T~Tauri stars \citep{Guedel06}.  From this perspective,
the X-ray emission of young brown dwarfs could be expected to be about
a factor of four below the level observed in older stars of the same
spectral type. From the X-ray observations of other young brown
dwarfs, one would expect a magnetic field of 0.5--1\,kG if X-ray
emission is directly proportional to magnetic flux (an admittedly
dubious proposition). Magnetic flux of a few hundred Gauss would still
be consistent with our result.

\section{Conclusions}

Our main conclusion is that current theoretical predictions of
magnetic fields required for funneled accretion are consistent with
the first upper limit on the field strength in young brown dwarfs,
i.e., with magnetic field strengths no larger than a few hundred
Gauss. This implies that accretion in low mass objects requires much
lower magnetic field strengths than in more massive stars, and that
the actual field strength may not grow much beyond what is needed to
regulate angular momentum during this part of brown dwarf evolution.
Along with the regulation of angular momentum by stellar magnetic
fields via disk-locking, this is a hint that the stellar fields may be
regulated by disk accretion in return.

We find no magnetic fields in four accreting young brown dwarfs, but
we detect a strong magnetic field in the non-accreting young low-mass
star UpSco~55. The two differences between UpSco~55 and the other four
targets are that (1) UpSco~55 has a mass above the hydrogen-burning
limit, and (2) that it shows no accretion. Older stars (without
accretion) above the hydrogen-burning limit but at similar temperature
also show very strong fields. Thus, two ways emerge to explain the
lack of magnetic flux in young brown dwarfs. First, magnetic fields
may be somehow regulated by disk accretion, or, second, that the
generation of magnetic flux is less effective in (young) brown dwarfs.
In either case, the usual dependence on Rossby number seems to be
violated, and apart from bulk rotation, it is hard to see how the
presence of a disk could influence production of fields deep inside a
star.

\citet{JK07} finds that the situation may be complicated in part by
how much of the total field is in a dipole component, but we appear to
be seeing a general qualitative dependence between the presence of
accretion and stellar magnetic flux. This should remain in place even
after accretion has ended, until the object begins to contract and
spin up. It will be important to make further direct field
measurements on young brown dwarfs, both accreting and non-accreting,
and also in old brown dwarfs.

It would have been easy (and not unexpected) for us to have found much
stronger fields on our targets. To remain consistent with accretion
regulation of angular momentum, however, we should then also have
found a much lower rotation velocity. It is of interest, therefore,
that to the extent we have been able to ascertain, in these fully
convective objects there is a consistency of surface rotation
velocities, and production of magnetic heating near the typical
saturation limit, over a wide range of stellar ages, masses, and
luminosities. This consistency is produced while balancing rather
different accretion rates, truncation radii, rotation periods, and
magnetic field strengths for different masses in young objects. It
will be informative to follow these hints on the ways in which initial
stellar parameters are tied to the conditions of star formation.

%% In a manner similar to \objectname authors can provide links to dataset
%% hosted at participating data centers via the \dataset{} command.  The
%% second curly bracket argument is printed in the text while the first
%% parentheses argument serves as the valid data set identifier.  Large
%% lists of data set are best provided in a table (see Table 3 for an example).
%% Valid data set identifiers should be obtained from the data center that
%% is currently hosting the data.

\acknowledgements

Based on observations collected at the European Southern Observatory,
Paranal, Chile, PID 279.C-5026, taken from the ESO Science Archive
Facility, PID 067.C-0160, and observed from the W.M. Keck Observatory,
which is operated as a scientific partnership among the California
Institute of Technology, the University of California and the National
Aeronautics and Space Administration. We would like to acknowledge the
great cultural significance of Mauna Kea for native Hawaiians and
express our gratitude for permission to observe from atop this
mountain.  A.R. has received research funding from the DFG as an Emmy
Noether fellow (RE 1664/4-1).  G.B. thanks the NSF for grant support
through AST00-98468.

\end{document}